# Realization of an acoustic third-order topological insulator

Haoran Xue[1], Yahui Yang[1], Guigeng Liu[1], Fei Gao[2], Yidong Chong[1,3,*] and Baile Zhang[1,3,*]

[1]Division of Physics and Applied Physics, School of Physical and Mathematical Sciences, Nanyang Technological University, Singapore 637371, Singapore.

[2]State Key Laboratory of Modern Optical Instrumentation, and College of Information Science and Electronic Engineering, Zhejiang University, Hangzhou 310027, China.

[3]Centre for Disruptive Photonic Technologies, Nanyang Technological University, Singapore 637371, Singapore.

*Author to whom correspondence should be addressed; E-mail: yidong@ntu.edu.sg (Y. Chong); blzhang@ntu.edu.sg (B. Zhang)

**The recent discovery of higher-order topological insulators (TIs) [1-5] has opened new possibilities in the search for novel topological materials and metamaterials. Second-order TIs have been implemented in two-dimensional (2D) systems [6-19] exhibiting topological 'corner states', as well as three-dimensional (3D) systems having one-dimensional (1D) topological 'hinge states' [20]. Third-order TIs, which have topological states three dimensions lower than the bulk (which must thus be 3D or higher), have not yet been reported. Here, we describe the realization of a third-order TI in an anisotropic diamond-lattice acoustic metamaterial. The bulk acoustic bandstructure has nontrivial topology characterized by quantized Wannier centers. By direct acoustic measurement, we observe corner states at two corners of a rhombohedron-like structure, as predicted by the quantized Wannier centers. This work extends topological corner states from 2D to 3D, and may find applications in novel acoustic devices.**

Higher-order TIs are a new class of topological materials supporting a generalization of the bulk-boundary correspondence principle, in which topological states are guaranteed to exist along boundaries two or more dimensions smaller than that of the bulk [1-5]. In standard TIs, topological edge states occur at one lower dimension than the bulk [21,22]; for instance, a quantum Hall insulator has a 2D bulk and topological states on 1D edges. By contrast, a 2D second-order TI supports zero-dimensional (0D) topological 'corner states'. Such a lattice was first devised based on quantized quadrupole moments [1,2] and quickly realized in mechanical [6], electromagnetic [7], and electrical [9] metamaterials. Later, another type of 2D second-order TIs based on quantized Wannier centers, was proposed [23-25] and demonstrated in acoustic metamaterials [10,11]. In 3D materials, second-order TI behavior has also been observed in the form of 1D topological 'hinge states' in bismuth [20].

According to theoretical predictions, TIs of arbitrarily high order are possible. However, in real materials the bulk is at most 3D. Thus, barring the use of 'synthetic' dimensions [26,27], the only remaining class of high-order TI is a *third*-order TI with 3D bulk and 0D corner states. As of this writing, no such material has been reported in the literature, although there exists a theoretical proposal based on quantized octupole moments [1,2].

Here, we realize a third-order TI in a 3D acoustic metamaterial, observing topological states at the corners of a rhombohedron-like sample. This third-order TI is based on the extension of Wannier-type second-order TIs to 3D [23,25], and can be regarded as a 3D generalization of the classic 1D Su-Schrieffer-Heeger (SSH) model [28]. Just as in the SSH case, the eigenmode polarizations are quantized by lattice symmetries, and the Wannier centers are pinned to high-symmetry points; the mismatch between the Wannier centers and lattice truncations gives rise to charge fractionalization and hence lower-dimensional topological boundary states [23,25]. This mechanism has previously been used to implement second-order TIs in acoustic kagome lattices [10,11].

The acoustic metamaterial is based on an anisotropic diamond lattice, with cubic unit cell shown in Fig. 1(a). The lattice constant is $a/\sqrt{2}$ and the three primitive lattice vectors are $a_1 = (a/2, a/2, 0)$, $a_2 = (0, a/2, a/2)$, $a_3 = (a/2, 0, a/2)$, where $a$ is the side length of cubic cell. The two sublattice atoms are located at $(0, 0, 0)$ and $(a/4, a/4, a/4)$. There are two sets of nearest-neighbor couplings. The couplings along [111] (plotted in red in Fig. 1(a)) have strength $t_2$, and those along other directions (plotted in blue) have strength $t_1$. When $|t_1/t_2|<1/3$, the lattice is a higher-order TI of the Wannier type [25]. The Wannier center is located at $(P_x, P_y, P_z)$, where the polarization along each direction is given by $P_i = -\frac{1}{V}\int_{BZ} A_i d^3k$, with V being the volume of the first Brillouin zone and $A_i = -i<\psi|\partial_{k_i}|\psi>$ being the Berry connection of the lowest band. For $|t_1/t_2|<1/3$, the Wannier centers are fixed at (1/2, 1/2, 1/2), which correspond to the centers of the red bonds in Fig. 1(a) (see Supplementary Information A for tight-binding calculations). Thus, when cutting through these bonds to form a finite sample, fractional charges will reside at the boundaries, similar to the SSH chain [28] and its previously-studied 2D generalizations [23-25]. In the present 3D case, we use this principle to predict and verify the existence of topological states at certain corners. The distribution of Wannier centers also allows us to explain the existence of surface states on certain 2D surfaces of the sample.

We assemble the lattice using coupled acoustic resonators [29-33], as shown in Fig. 1(b). The two identical thick cylindrical resonators correspond to the two sublattice atoms in Fig. 1(a), which are connected and coupled by the other thinner cylindrical waveguides. The entire structure is filled with air, and the boundary walls are regarded as hard boundaries. Here the three lattice vectors are the same as in the tight-binding model in Fig. 1a, with $a = 175$ mm. The height ($H$) and radius ($r$) of each cylindrical resonator are 60 mm and 20 mm, respectively. For these parameters, a single cylindrical resonator has a fundamental mode at 2883.3 Hz (see Fig. 1(c) for the mode profile). The two sets of coupling strength are realized by tuning the radius of the connecting waveguides. The larger connecting waveguides, with radius $r_{c2}$, correspond to $t_2$ bonds in Fig. 1(a); the others, with radius $r_{c1}$, correspond to $t_1$ bonds. All connecting waveguides are located at $h = 8.125$ mm, measured from either the top or bottom of each resonator, as indicated in Fig. 1(b).

When $r_{c1} = r_{c2}$ (i.e., $t_1 = t_2$), the bulk bandstructure exhibits nodal lines (Fig. 1(e), red curve; see Fig. 1(d) for the 3D Brillouin zone) [25,34]. The higher-order TI phase ($|t_1/t_2|<1/3$) is achieved by tuning the radii of the connecting waveguides. The blue curve in Fig. 1(e) shows the simulated bulk bandstructure for $r_{c1} = 2$ mm and $r_{c2} = 7.8$ mm, which has a bandgap between the two bands.

By fitting simulated dispersions with the tight-binding calculation, the ratio $t_1/t_2$ can be estimated to be around 0.076 (see Supplementary Information A for details), indicating that the system is a higher-order TI.

A third-order TI is characterized by the existence of topological corner states at certain corners of a finite 3D sample. We construct a rhombohedron-like structure (Fig. 2(a)) containing six rhombus-shaped surfaces (see Fig. 2(b) for the ($\bar{1}\bar{1}1$) surface). Each surface constitutes a finite anisotropic honeycomb lattice, and can be regarded as a second-order TI [25]. As shown in Ref. 25, corner states in this rhombohedron-like structure can be found at two corner resonators, labeled 'A' and 'B' in Fig. 2(a). The Wannier centers are located at the centers of the red bonds in this case. As can be seen from Fig. 2(a), in this rhombohedron-like structure all atoms are connected to red bonds except for the two corner atoms at 'A' and 'B', where corner states are expected to appear. To check this prediction, we perform acoustic simulations on a structure containing 52 resonators (see Figs. 2(d)-(e) for the structure). As shown in Fig. 2(c), there exist two in-gap modes at around 2,891 Hz, between the upper bulk band around 3,000 Hz and the lower bulk band around 2,800 Hz. The eigenmode patterns, shown in Figs. 2(d)-(e), reveal that the acoustic pressure is highly concentrated on the two corners, thus verifying that these are indeed third-order topological boundary states.

To experimentally demonstrate these corner states, we fabricate a sample (see Fig. 2(f)) through stereo-lithography 3D printing, with the same parameters as in the preceding simulations. We then identify the extents of the upper and lower bulk bands by measuring bulk transmission through two resonators labeled 'C' and 'D' in Fig. 2(a). As shown in Fig. 3(a), the spectrum exhibits two clear peaks, corresponding to the upper and lower bulk bands. Next, we measure the local acoustic response at the two corner resonators, 'A' and 'B' (see Supplementary Information

D for details on the measurements). For both resonators, we observe a peak at around 2,900 Hz, corresponding to the corner states (see Supplementary Information B for discussions on robustness of the corner states). Finally, we repeat the acoustic response measurement for all resonators, constructing an intensity map of the lattice equivalent to the local density of states. The results of these measurements, conducted at the peak frequency of the corner resonances (2,900 Hz), are plotted in Figs. 3(c)-(d) from different view angles. We observe that the acoustic pressure is indeed highly concentrated at the two relevant corners: there is negligible response at the bulk sites, along the 2D surfaces or 1D corners, or at sample corners not corresponding to divided Wannier centers.

It is important to note that no 2D or 1D surface states are observed in this sample because the lattice is truncated in such a manner that *only* the two corner resonators are 'peeled off' from the red bonds (see Fig. 2(a)). However, it is also possible to truncate the lattice in other ways that should generate topological surface states, according to the analysis of quantized Wannier centers. Previously, it has been predicted that such surface states should exist on the (111) surface of such a lattice [34] (see Supplementary Information C); moreover, there should be no surface states along the other surfaces where the lattice truncation occurs along blue bonds. To test this prediction, we fabricated a tetrahedron-like sample containing four different surfaces, oriented at $(111)$, $(1\bar{1}\bar{1})$, $(\bar{1}1\bar{1})$ and $(\bar{1}\bar{1}1)$, as shown in Fig. 4(a). We excite and measure the acoustic response at a surface resonator on the (111) surface (indicated by the red star in Fig. 4(a)), and then repeat the procedure for the other three surfaces. The results, shown in Fig. 4(b), agree well with the theoretical prediction. Only along the (111) surface is there a sharp response peak, at around 2910 Hz, corresponding to the surface state; along the other three surfaces, we observe only the two peaks corresponding to the bulk states. By performing the measurement in all the lattice resonators, we derive the map shown in Figs. 4(c)-(d): at 2900Hz, the response is much higher along the (111) surface than elsewhere in the lattice.

In conclusion, we have implemented a third-order TI on an acoustic anisotropic diamond lattice. Corner states and surface states were observed in a rhombohedron-like sample and a tetrahedron-like sample, respectively, in accordance with a theoretical analysis based on quantized Wannier centers. The extension of topological corner states from 2D to 3D may have potential use in applications such as acoustic manipulation and sensing [35]. We also envision that our study will inspire more experimental studies into the implementation of higher-order TIs in higher dimensions.

## Acknowledgements

The authors thank Wenzheng Ye for help on sample fabrication. This work was sponsored by Singapore Ministry of Education under Grants No. MOE2015-T2-1-070, MOE2015-T2-2-008, MOE2016-T3-1-006 and Tier 1 RG174/16 (S).

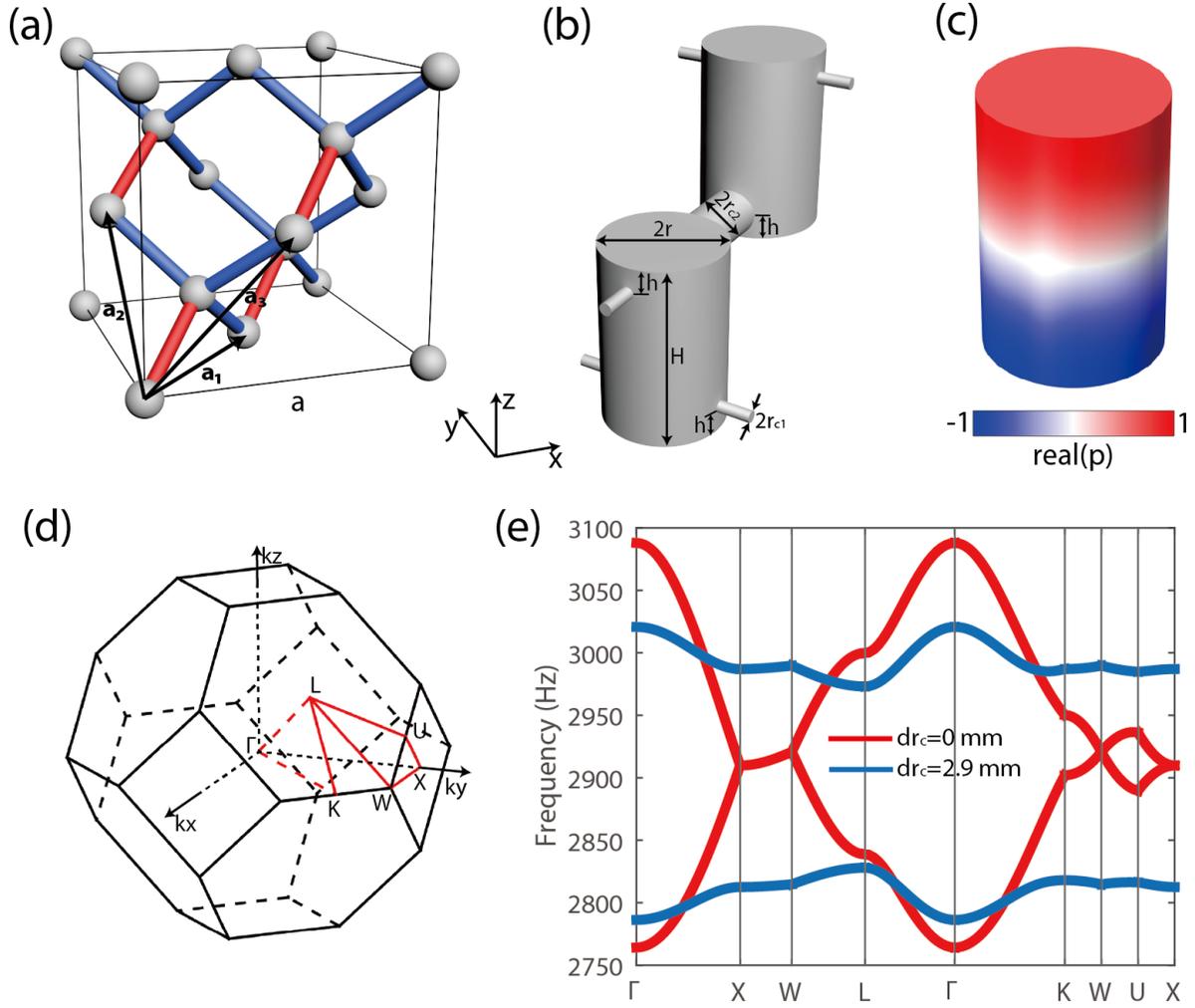

FIG. 1. Structure and bulk dispersion of acoustic anisotropic diamond lattice. (a) Schematic of the cubic cell of an anisotropic diamond lattice. Red (blue) bonds indicate couplings with strength $t_{2(1)}$. (b) The unit cell of the acoustic anisotropic diamond lattice. All parameter values are given in the main text. (c) Acoustic pressure profile for the resonator eigenmode of interest in this work. (d) The first Brillouin zone of the anisotropic diamond lattice. (e) Simulated bulk bands of acoustic anisotropic diamond lattice shown in (b). Red curve: $dr_c = 0$ mm ($r_{c1} = r_{c2} = 4.9$ mm); Blue curve: $dr_c = 2.9$ mm ($r_{c1} = 2$ mm and $r_{c2} = 7.8$ mm).

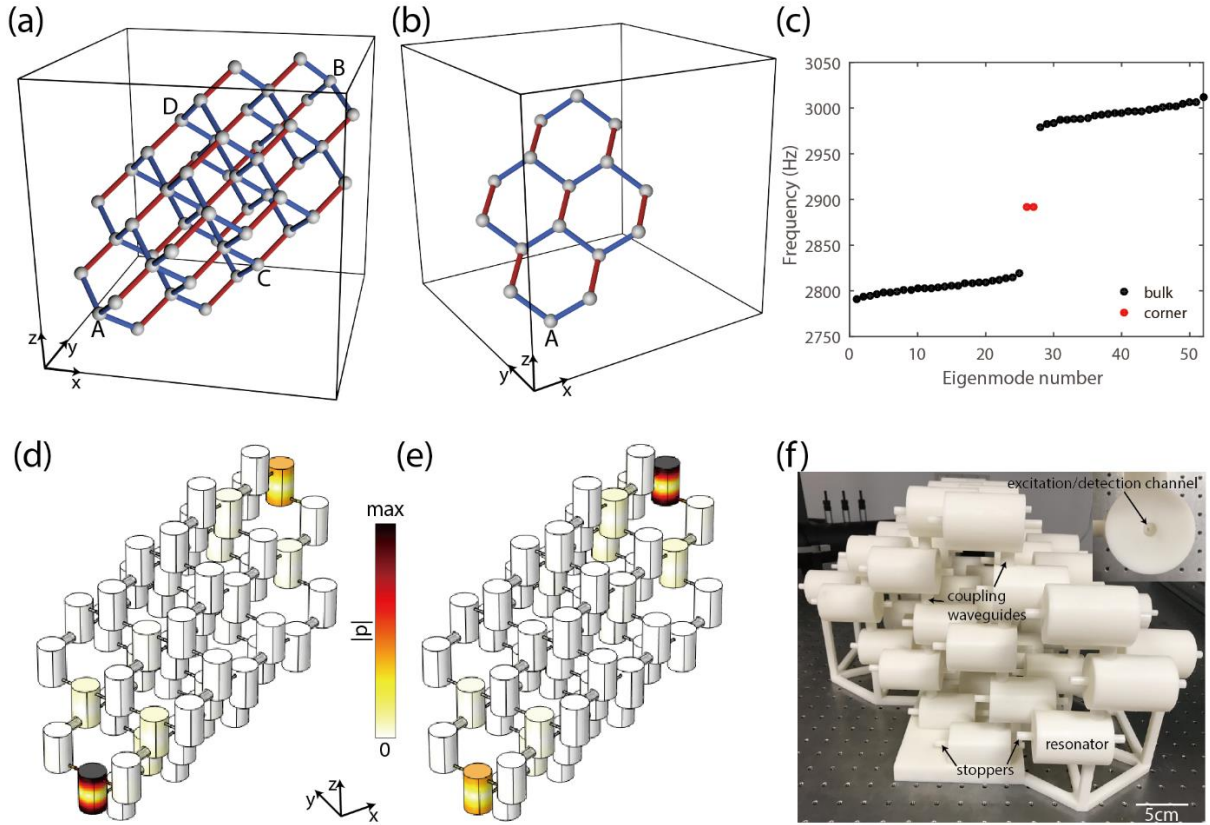

FIG. 2. Topological corner states on a rhombohedron-like sample. (a) Schematic of the rhombohedron-like structure. (b) ($\bar{1}\bar{1}1$) surface atoms of the structure shown in (a). (c) Simulated eigenfrequencies of the finite acoustic lattice. Black and red dots represent bulk and corner states, respectively. (d) and (e) Eigenmode profiles of the two corner states. (f) Photograph of the fabricated rhombohedron-like sample containing 52 resonators. The inset shows a zoomed-in view of a resonator with one stopper removed.

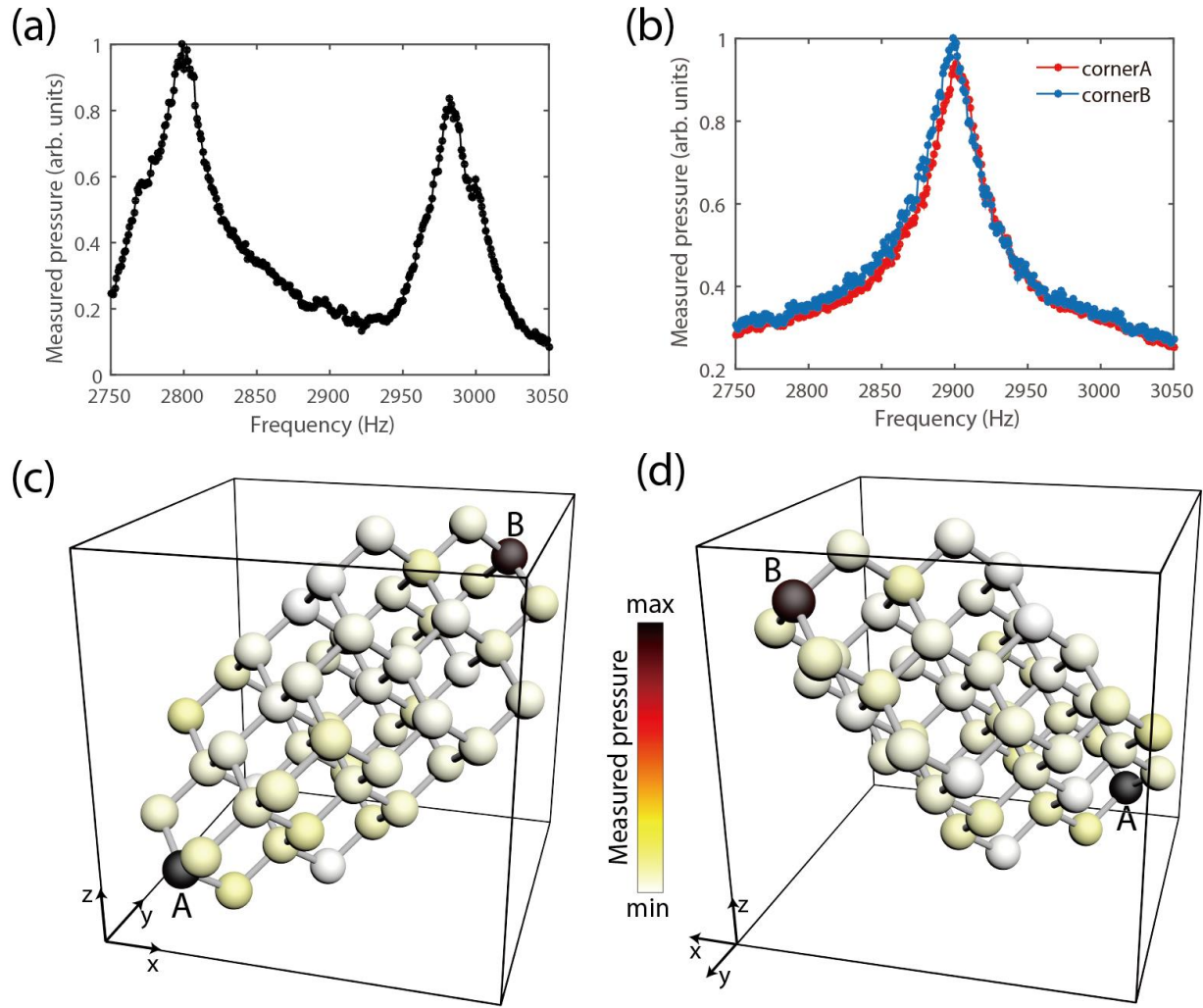

FIG. 3. Experimental observation of corner states in a rhombohedron-like acoustic structure. (a) Measured bulk transmission spectrum. (b) Measured spectra at the two corners (positions located in (c) and (d)). (c) and (d) Measured spatial map at 2,900Hz. (c): front view; (d): back view. The structure is simplified for the purposes of illustration; for the real structure see Figs. 2(d)-(f). The balls correspond to the cylindrical resonators and the gray bonds indicate the couplings. The color of each ball represents the measured acoustic pressure at that site, as measured at the top of the cylindrical resonator. The colormap does not apply to the bonds.

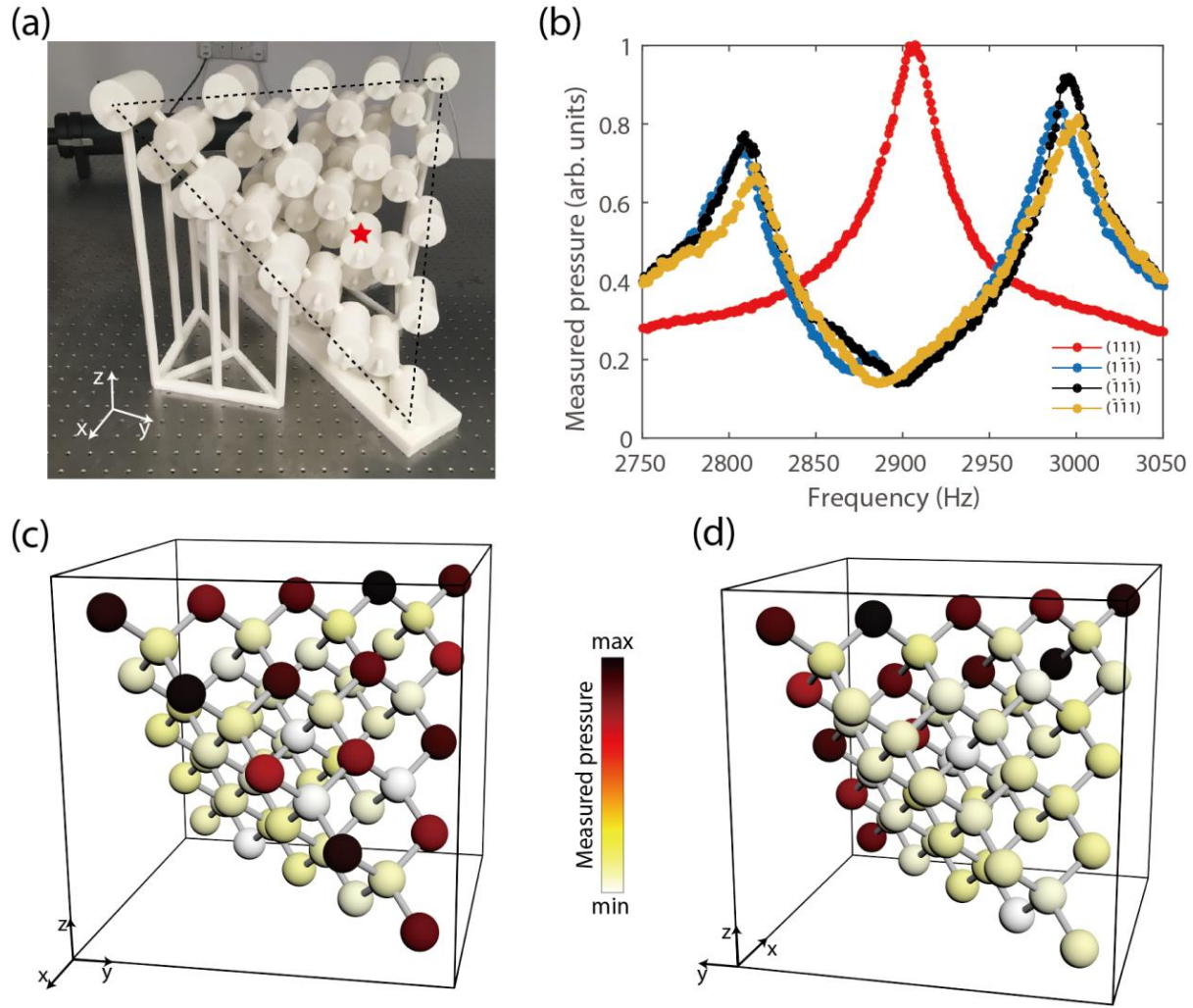

FIG. 4. Experimental demonstrations of surface states on a tetrahedron-like acoustic structure. (a) Photograph of the fabricated sample. Black dashed line denotes the (111) surface. (b) Measured spectra at four surface resonators on four surfaces of the sample. The source and probe are located at the same resonator for each curve. (c) and (d) Measured spatial map at 2,900 Hz. (c): front view; (d): back view.